# Bi-phase age-related brain gray matter magnetic resonance T1ρ relaxation time change in adults

1* Yáo T Li *BSc*, 2* Hua Huang *MMed*, 3 Zhizheng Zhuo *PhD*, 2 Pu-Xuan Lu *MB*, 1 Weitian Chen *PhD*, and 1 Yì Xiáng J Wáng *MMed*, *PhD*

1. Department of Imaging and Interventional Radiology, Faculty of Medicine, The Chinese University of Hong Kong, Prince of Wales Hospital, Shatin, New Territories, Hong Kong SAR
2. Department of Radiology, Shenzhen No. 3 People's Hospital, Shenzhen, Guangdong Province, People's Republic of China.
3. MR Clinical Sciences, Philips Healthcare Greater China, Beijing, People's Republic of China.

* These two authors contributed equally to the work.

Address correspondence to: Dr Yì Xiáng J Wáng
E-mail: yixiang_wang@cuhk.edu.hk

Running title:
Age-related brain T1ρ change

**Abstract**

**Objectives:** To investigate normative value and age-related change of brain magnetic resonance T1ρ relaxation at 1.5T.

**Methods:** 20 males (age: 40.7±15.5 years, range: 22-68years) and 22 females (age: 38.5 ± 14.8 years, range: 21-62 years), were scanned at 1.5 Tesla using 3D fluid suppressed turbo spin echo sequence. Regions-of-interests (ROIs) were obtained by atlas-based tissue segmentation and T1ρ was calculated by fitting the mean value to mono-exponential model. Correlation between T1ρ relaxation of brain gray matter regions and age was investigated.

**Results:**   A regional difference among individual gray matter areas was noted; with hippocampus (98.37±5.37 msec) and amygdala (94.95±4.34 msec) have the highest measurement, while pallidum (83.81±5.49) and putamen (83.93+4.76) the lowest measurement. T1ρ values decreased slowly (mean slope: -0.256) and significantly ($p<0.05$) with age in gray matter for subjects younger than 40 years old, while for subjects older than 40 years old there was no significant correlation between T1ρ relaxation and age.

**Conclusion:** T1ρ relaxation demonstrates a bi-phase change with age in adults of 22-68 years.

## Introduction

Matter magnetic T1ρ relaxation has the potential to provide information about the low frequency motions (100Hz to a few kilohertz) in biological systems [1, 2]. T1ρ relaxation not only depends on T1 and T2 but also has contributions from several MR interactions such as chemical exchange, dipolar interaction and J-coupling. Depending upon the tissue type, more than one mechanism may be operative simultaneously but with different relative contributions. During recent years, T1ρ relaxation has been increasingly used to explore the pathophysiology or predictive diagnostics of a number of neurological conditions [4-10]. Previous studies suggested that neurodegeneration may contribute to the increased T1ρ in brain regions [11-13]. For example, cerebral atrophy, which reflects underlying neuronal loss, has been reported to be associated with the increased T1ρ values in the hippocampus and medial temporal lobe of Alzheimer's disease [11]. Although the biophysical/biochemical mechanism remains to be further investigated, novel MRI techniques such as T1ρ, T1ρ dispersion, chemical exchange saturation transfer and its variant chemical exchange imaging with spin-lock technique may provide early imaging biomarkers for neurodegeneration diseases including Alzheimer's disease, Parkinson's disease and dementia [14-18]. These techniques have been refined to be increasingly more time-efficient, faster and more robust in recently years [19-30].

T2 relaxation has also been explored as one of the contrast mechanism in disease characterization [31, 32]. The relationship between T2 relaxation and physiological ageing remains not confirmed yet. Siemonsen *et al* [33] studied 50 subjects (age: 12–91 years) and found an increase in T2 that linearly correlated with age in the thalamus and three white matter structures, but not in the caudate nucleus and lentiform nucleus. Ding *et al* [34] studied 70 normal subjects (age: 3 weeks-31 years) and showed that T2 decreased with increasing age; the rate of decrease was greater at a younger age and slower in the years after, indicating a nonlinear relationship with age. Hasan *et al* [35] studied 130 healthy subjects (age: 15–59 years) and reported the relation between T2 and age in whole brain gray and white matter, caudate nucleus, and the anterior limb of internal capsule followed a quadratic, U-shaped curve. More recently, Wang *et al.* [36] studied 77 normal subjects (age: 9-85 years) and reported brain tissue R2 (1/T2)-age correlations followed various time courses with both linear and nonlinear characteristics depending on the particular brain structure.

T1$\rho$-age relationship has not been studied as much as T2 relaxation. Borthakur *et al* [37] found no relationship between T1$\rho$ vs. age in 16 elderly subjects (age: 70-91 years). Recently Watts *et al* [38] studied 41 subjects (age: 18-76 years) and reported T1$\rho$ values significantly decrease with age in cortical gray matter, left and right caudate, putamen, hippocampus, amygdala, and nucleus accumbens; while increases with age were observed in white matter tracts. Information on age-related change in T1$\rho$ relaxation is not only needed to gain a deeper understanding of brain ageing but also useful as a normative data set for examinations performed with T1$\rho$ MRI in patients. The primary goal of this study was to further clarify the discrepancy seen in the reported literature [37, 38].

## Materials and Methods

Subjects

This study was approved by the local ethical committee. There were 42 adults volunteers, including 20 males (ages 22-68 years, with a mean and standard deviation of 41± 16 years) and 22 females (ages 21–62 years, with a mean and standard deviation of 39 ± 15 years). None of the subjects had neurological diseases except those generally associated with ageing in older subjects [39].

MRI protocol

The MRI data sets were acquired with a 1.5 T Philips Achieva scanner (Philips Healthcare, Best, the Netherlands) using a 16 channel head coil (Invivo Corp, Gainesville, USA). The pulse sequence was similar to that used in [38]. The T1ρ-weighted images were acquired using 3D fluid suppressed turbo spin echo (TSE) sequence. A combination of T2-preparation and inversion recovery was used for fluid suppression, which was equivalent to the fluid suppression method reported by Wong *et al* [40]. Adiabatic refocusing and inversion pulses were used for T2-preparatin and the $180^0$ inversion respectively, to improve the robustness against $B_1$ inhomogeneity. The TI (time of inversion) used for fluid suppression was 1650msec. The spin-lock was achieved using a rotary echo approach as described by Charagundla *et al* [41]. Six T1ρ-weighted images were acquired with total spin lock time (TSL) 0, 20, 40, 60, 80, and 100msec. After T1ρ-prep, the imaging sequence consists of a 3D TSE sequence with variable flip angle. The imaging parameters include: sagittal plane, FOV 25cm x 25cm, TR/TE 4800/228ms, 3D isotropic resolution $1.8 \times 1.8 \times 1.8 \text{mm}^3$, 100 slices with whole brain coverage, TSE factor 182 with echo spacing 2.3msec, SENSE acceleration factor 2.6 along anterior-posterior and 2 along right-left direction, and NSA 2. The acquisition time of each spin lock time was 1 min 50 s, giving a total acquisition time about 7min 30sec.

Data Analysis

In order to achieve the brain motion-correction for all the images with different TSLs caused by involuntary brain movement, respiration and CSF pulsation, all the T1ρ images with different TSLs were realigned to the first T1ρ image (TSL=0). The individual T1ρ image with TSL=0 was segmented into gray matter and white matter by using a series of Tissue Probability Maps (TPM) templates in MNI Montreal Neurological Institute space using SPM8 (Functional Imaging Laboratory, The Wellcome Trust Center for NeuroImaging, Institute of Neurology, UCL University College London, London, UK) [42]. Tissue Probability Maps assign voxels a probability of belonging to gray matter or white matter or CSF or other tissues. The subject-specific segmented gray matter and white matter images in original T1ρ image (native space) were gray matter (GM) and white matter (WM) distribution probability mapping with a range of 0-1. The individual gray and white mater binary masks, i.e. images with only 0 and 1 values, were obtained by thresholding the segmented images with a threshold of 0.5. For acquiring gray matter binary masks, the voxels was defined belonging to gray matter if the probability was higher than 0.5, and voxel value was set to 1 if probability was higher than 0.5 and otherwise voxel value was set to 0; and the same was applied to white matter binary masks [38, 43-45].

The individual gray and white matter binary masks were applied on the original T1ρ images with different TSLs to extract the gray matter and white matter [38]. Optimized atlas-based methods provides more accurate and specific diagnostic surrogate markers compared to whole brain histograms, voxel-by-voxel or large ROI methods. Specific region-of-interests (ROIs) such as hippocampus were extracted based on the Anatomical Automatic Labeling template (in MNI Montreal Neurological Institute space) by using REST software (Resting-State fMRI Data Analysis Toolkit, State Key Laboratory of Cognitive Neuroscience and Learning, Beijing Normal University, Beijing, China) [46]. All the original T1ρ images were normalized to the MNI space by using the normalization parameters generated in New Segment batch in SPM8. In this way, all the T1ρ images were warped into the MNI space same as ROI masks. Finally, the ROI masks were applied on the normalized T1ρ images to calculate the T1ρ values within specific ROIs. The segmentation results were visually verified by an experienced radiologist.

The T1ρ-weighted images are assumed to follow a mono-exponential decay model:

$$I(SLT) = I_0 \times \exp(-SLT/T1\rho) \qquad [1]$$

Where $I_0$ is the image when SLT is zero. To avoid potential measurement bias due to relative lower SNR at 1.5T (supplement document 1), after signal averaging within ROI, non-linear least square fit with the Levenberg-Marquardt algorithm was used to fit the data to the mono-exponential model. All image analyses were conducted in Matlab (MathWorks Inc, USA). Weighted least square was used to access fitting quality. When poor fitting data ($R^2<0.95$) occurred, possible error was check again. Points with value > (mean +2SD) or < (mean-2SD) were discarded.

Statistical Analysis

All statistical analyses were performed with software SPSS22 (IBM Corp, Chicago, USA). Two-tailed Student t-test, which was validated using a Kolmogorov-Smirnov test for normality, was used to check the difference between T1ρ values of male and female, left and right brain. Despite the generally existing cerebral asymmetry [47], no asymmetry in MR relaxivity asymmetry has been reported between the left and right sides of the brain as well as no difference between males and females [36, 38, 40, 49], and our data analysis showed the same. Therefore the T1ρ values from two hemispheres were averaged for analysis.

Based on the initial visual inspection of the trend of T1ρ vs. age, and our polynomial fitting showed the lowest T1ρ value during the age period 43 to 54 years (supplement document 2), and plus with reference to data presented by Hasan *et al* [35] and Wang *et al* [36], the data in this study were divided into two groups, i.e. before 40 years group (n=22, 11 males and 11 females) and after 40 years group (n=20, 9 males and 11 females). Then linear regression and Pearson correlation coefficients between T1ρ values and age were calculated. Analysis of variance (ANOVA) with repeated measures and multiple comparisons were used to compare the T1ρ variation among different tissue regions.

**Results**

Table 1, 2 showed the mean, standard deviation (SD), Pearson correlation coefficient and p-value of gray matter and specific structures for subjects younger than 40 years and older than 40 years, respectively. In subjects younger than 40 years, T1ρ values in the selected gray matters showed significant negative correlation with age. However, for subjects older than 40 years, T1ρ values in the selected gray matters showed no significant correlation with age (Figure 1). Supplement document 3 shows if subjects of all age are grouped together for linear regression fit, the correlation is much reduced compared with younger subjects only (< 40 years).

Analysis of variance (ANOVA) with repeated measures showed T1ρ values varied in different gray matter structures (Figure 2), with Amygdala and Hippocampus had the highest mean T1ρ values of 94.95±4.34 msec and 98.37±5.37 msec, and putamen and pallidum had the lowest mean T1ρ values of 83.81±5.49 msec and 83.93+4.76msec.

Global white matter measured T1 ρ value of 88.65±3.47 msec, and the association with age was not significant (p=0.18, Supplement document 3).

**Discussion**

The diagnosis and therapy of age-related chronic diseases will become more and more challenging in the course of the next few decades because the percentage of elderly people is increasing. Objective and quantitative imaging strategies sensitive to early biochemical changes in brain tissue will benefit evaluation of potential new therapies and longitudinal monitoring of disease progression. In a pre-clinical study Plaschke *et al* [50] demonstrated MR Relaxometry may shows subtle changes not detected by neuropathological evaluations. If T1ρ imaging is to have application to neurodegenerative disease and other pathology, it is essential to understand its variation during normal aging and normative values for T1ρ across different ages.

T1ρ relaxation time constant is influenced by molecular processes that occur in the millisecond range, such as chemical exchange of protons between water associated with macromolecules and free water. In biological tissues, T1ρ is dependent on the macromolecular composition. Studies on the relationship between T1ρ and ageing are till now limited. Watts *et al*'s study at 3T estimated brain gray matter T1ρ values of 74 msec to 86 msec [36]. Further, Gonyea *et al* [4] reported normal appearing gray matter had T1ρ of 78.2 ± 1.3 msec. The gray matter mean T1ρ value measured in the current study ranged from 83.76 msec to 96.00 msec, which is longer than the data presented by Watts *et al* [36] and Gonyea *et al* [4]. On the other hand, the regional variation in our study is similar to Watts *et al* [36], where the amygdale and hippocampus also had the highest mean T1ρ values and putamen and pallidum had lowest mean T1ρ values, though this point was not specified in their report. One plausible explanation would be these regional T1ρ measurements were due to local iron deposition difference as discussed in Vymazal *et al*'s paper on T2 relaxivity [31]. Our measurements broadly agree with earlier reports at 1.5T. Borthakur *et al* [37] reported T1ρ values (1.5T) for gray matter of 86.4 ± 4.4 msec in their elderly control subjects (age 78 ±2 years). Haris *et al* [12] reported T1ρ values (1.5T) in the left and right hippocampus of 90.2 ± 12.3 and 92.2 ± 13.6 ms respectively in a control group aged 71.2 ± 9.8 years. T1ρ has higher value at 1.5T than at 3T supports the biophysics of T1ρ relaxation mechanism [1,2].

The T1ρ and T2 values are correlated with T2 can be regarded as a special case of T1ρ with a spin lock frequency of 0 Hz. It was reported at the spin lock frequency around 500 Hz, T1ρ provides additional sensitivity to low frequency processes [2, 4]. Recently, Wang *et al* [36] reported significant positive correlation between tissue R2 and age from 9 to 30 years old in gray matter structures; however, the R2-age correlation after age 40 demonstrated more diverse characteristics. Hasan *et al* [35] reported a quadratic relationship between global T2-values for gray matter with a minimum around age 45. Watts *et al* [38] reported a decrease of gray matter T1ρ with age. The inspection of our data suggested T1ρ of gray matter might follow a bi-phase change relative to age. The initial decreasing phase was till around 40 years old, while gray matter in subjects older than 40 years old showed no clear age relationship. We argue that a further look at Watts *et al's* data may also suggest there was no apparent trend in subjects older than 40 years as well (Fig 2 of reference 38). The decreasing trend was mainly contributed by subjects younger than 40 years [38], and the same might be true with Hasan *et al's* T2 relaxation results (Fig 5 of reference 35). Our results would also agree with the initial study of Borthakur *et al* [37], where no relationship was found between age and T1$\rho$ in elderly subjects. Note if a polynomial fitting approach is taken, the fitting-line on right-side is more likely showing a U-turn toward up than being fitted flat (supplementary Fig 2, Fig 5 of reference 35). Our study further demonstrates the sophisticated manner of the relationship of MR relaxation vs. age increase. It can be nonlinear, and also vary across brain regions [36]. The comforting side would be that no clear change associated with ageing was noted for subjects older than 40 years. We would like to suspect even more older subjects were recruited and a type of change pattern reaches statistical significance, this correlation will remain weak. Considering the physiological life span of human beings (~ 100 years), the biochemical process related to a decreasing T1$\rho$ value before 40 years old is more likely related to maturation than to ageing.

Recently, Zhao *et al* [51] reported a T1$\rho$ increase associated with ageing (~ 15 months) in rat brain gray matter areas of the thalamus, hippocampus and frontal cortical regions. This trend apparently differs from the results from the current study. However, Plaschke *et al* [50] also found a significant increase in T2 relaxation time in senescent (18-month-old) rats' brain compared with the same rats measuring at the age of 12 month, with the value for hippocampus increased from 77.2±29 to 110±28 msec (2.35-T B0 magnetic field). T2 relaxation time was specially increased after systemic hypotension treatment. Importantly, Plaschke *et al* showed increased T2 relaxation time in rat brain was negatively correlated with reference memory. In addition to biological differences between human brains and rodent brains and different relative ages, the bi-phase or a quadratic curve of brain tissue relaxivity changes may partially explain the T1$\rho$ divergence in this study and the results of Zhao *et al* [35, 36, 51].

There are a few limitations for this study. The number of study subject remains relatively small. Both children and very old subjects were not included in this study. The conclusion of age-dependant change in subjects older than 40 years cannot be firmly drawn from this study. The subjects were adjudged to be healthy based on that there were no specific neurological symptoms, and the MRI reading being normal except changes commonly seen in elderly subject [39]. We could not exclude the possibility that a few subjects had un-diagnosed psychiatry diseases. However, this would be a potential limitation for all similar studies, and it is unlikely this would change the conclusion of this study.

In conclusion, our study indicates the complexity of age-related T1ρ changes in the brain as the changes do not necessarily follow a linear pattern, and also can be structure specific. Careful observation suggests consistent patterns from multiple studies which have been published recently, with adults of aged <40 years demonstrating decreasing T1ρ relaxation correlated with age. Brain tissue T1ρ measurement at 1.5T is slightly longer than at 3T which supports the biophysics of T1ρ relaxation.

**Acknowledgement**:

We thank Queenie Chan PhD, Philips Healthcare Greater China, for her supports during the study.

**Fig Legends**

Figure 1: Brain gray matters T1ρ relaxation (Y-axis: msec.) vs. age ( X-axis: years). a: amygdala; b: caudate; c: cortices; d: hippocampus; e: pallidum; f: putamen; g: thalamus; h: global gray matter.

Figure 2.  Box-plot of T1ρ at various gray matter structures. A regional difference shows hippocampus and amygdala have the highest measurement, while pallidum and putamen the lowest measurement. Y-axis: T1ρ value at msec. * & [0]: outlier measurements.

Supplements

Supplement document 1: A comparison of ROI-based and voxel-wised Data Analysis

Supplement document 2:  Polynomial fitting for grey matter T1ρ value vs. age

Supplement document 3: Supplement Table 1. Descriptive Statistics, Correlation with Age for T1rho value of global WM, GM and selected gray matter structure with linear regression fit.

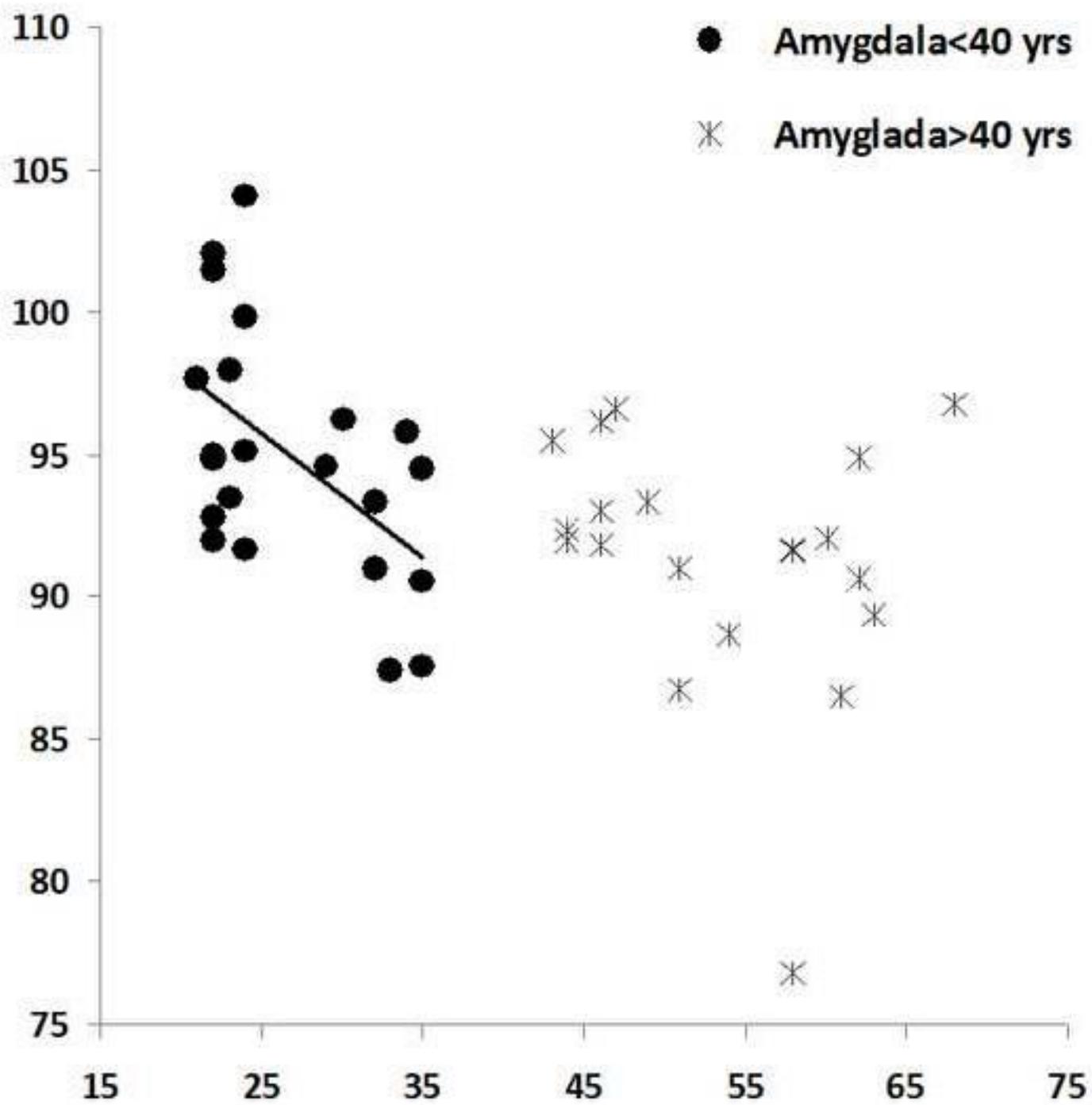

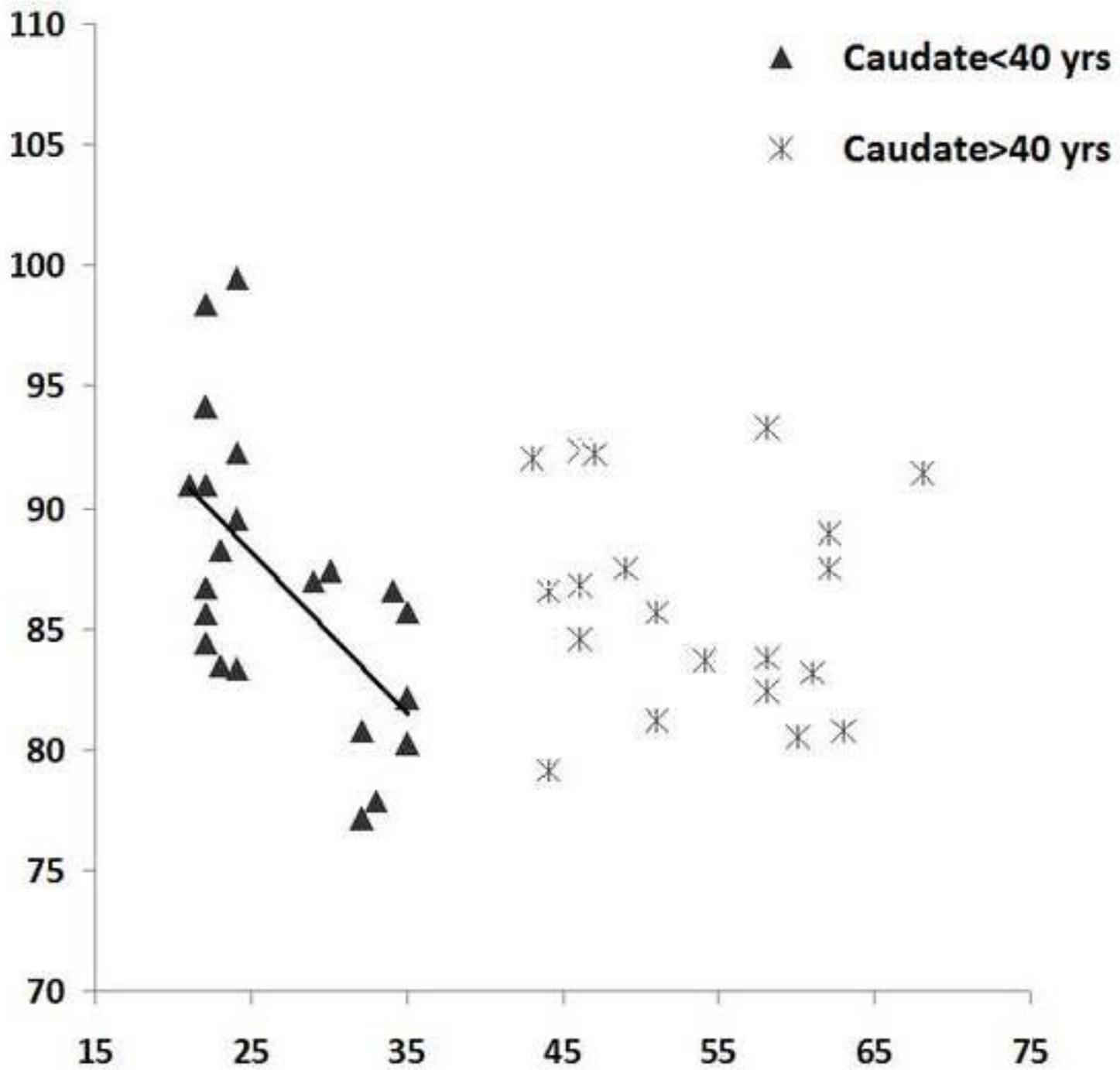

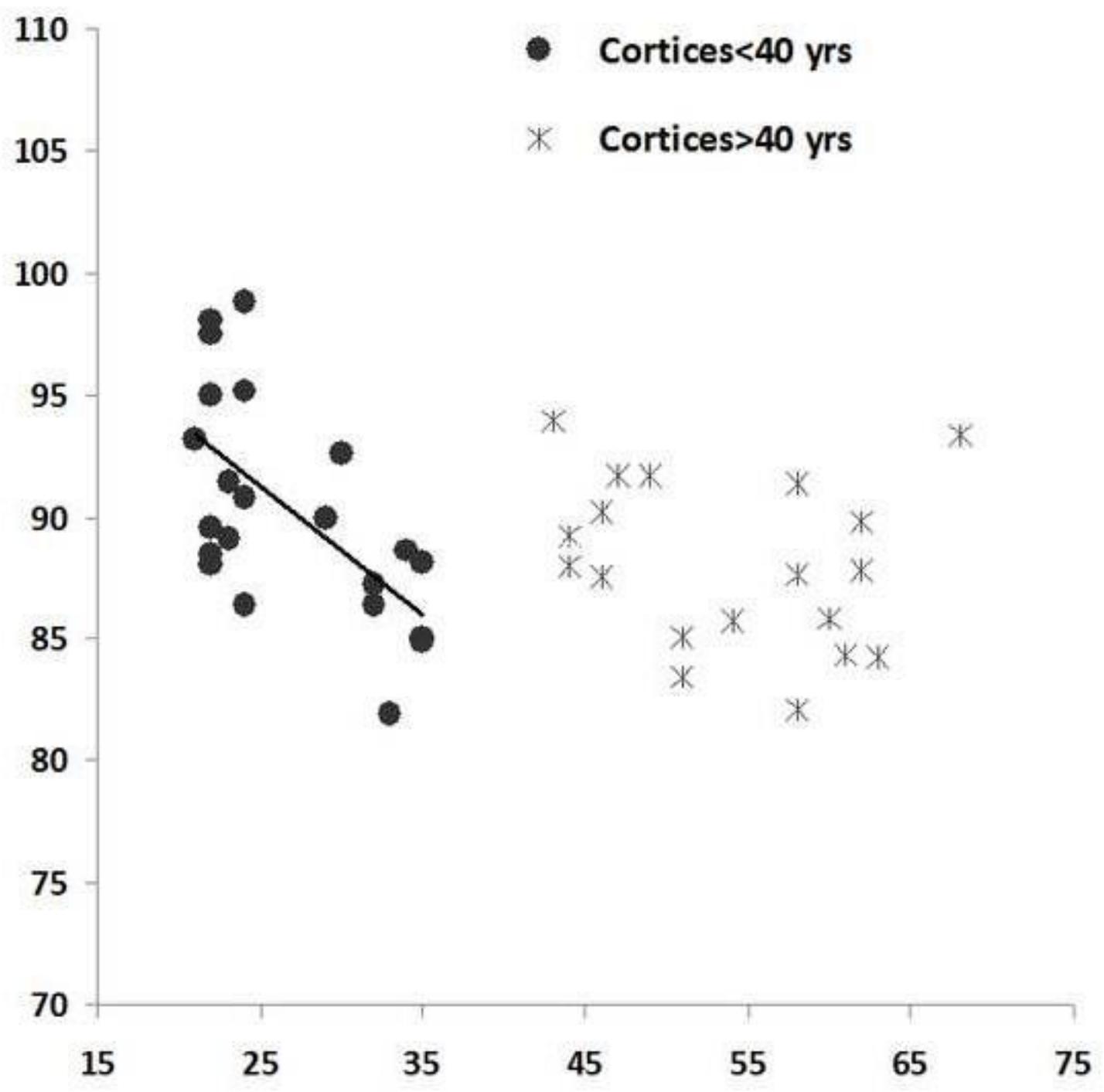

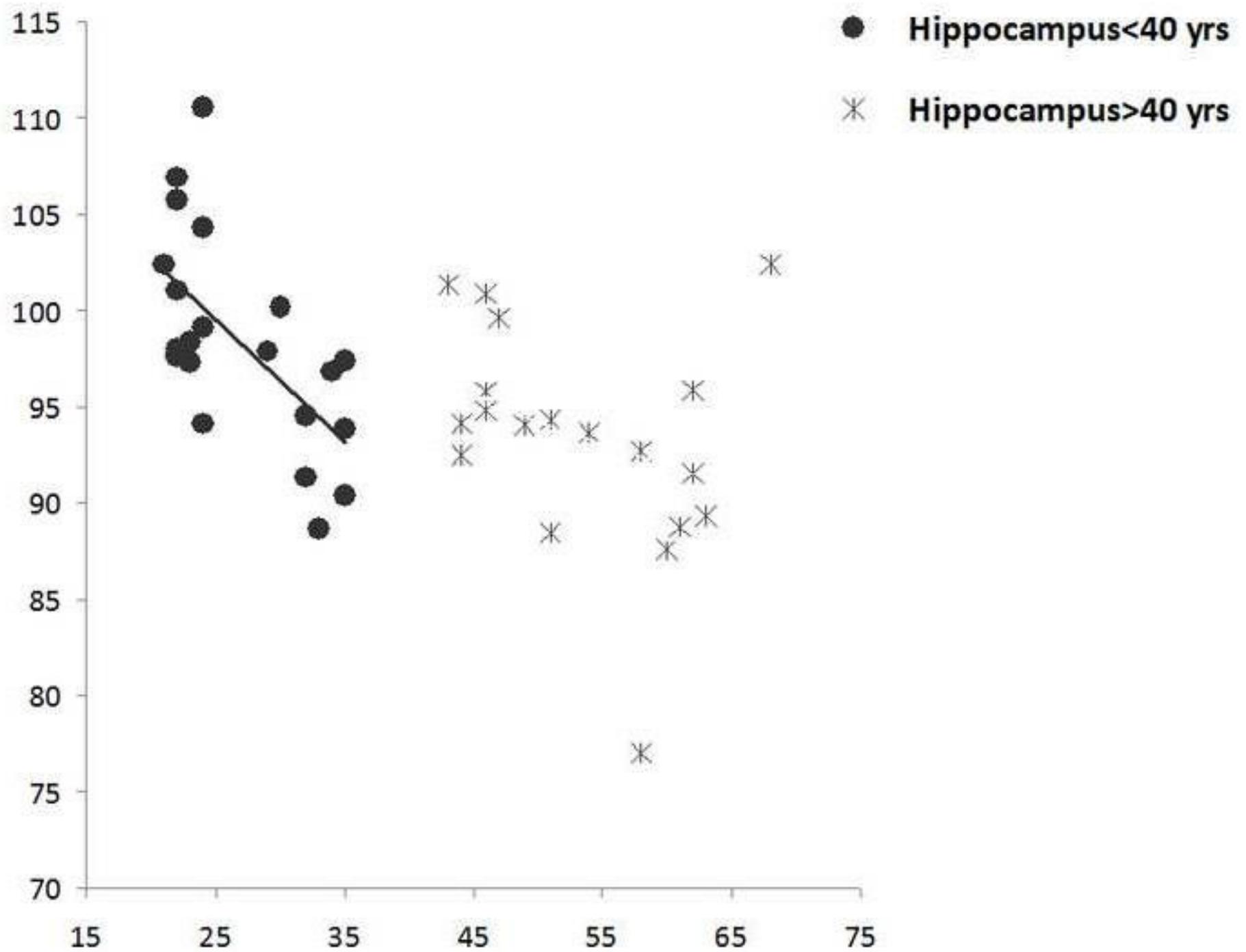

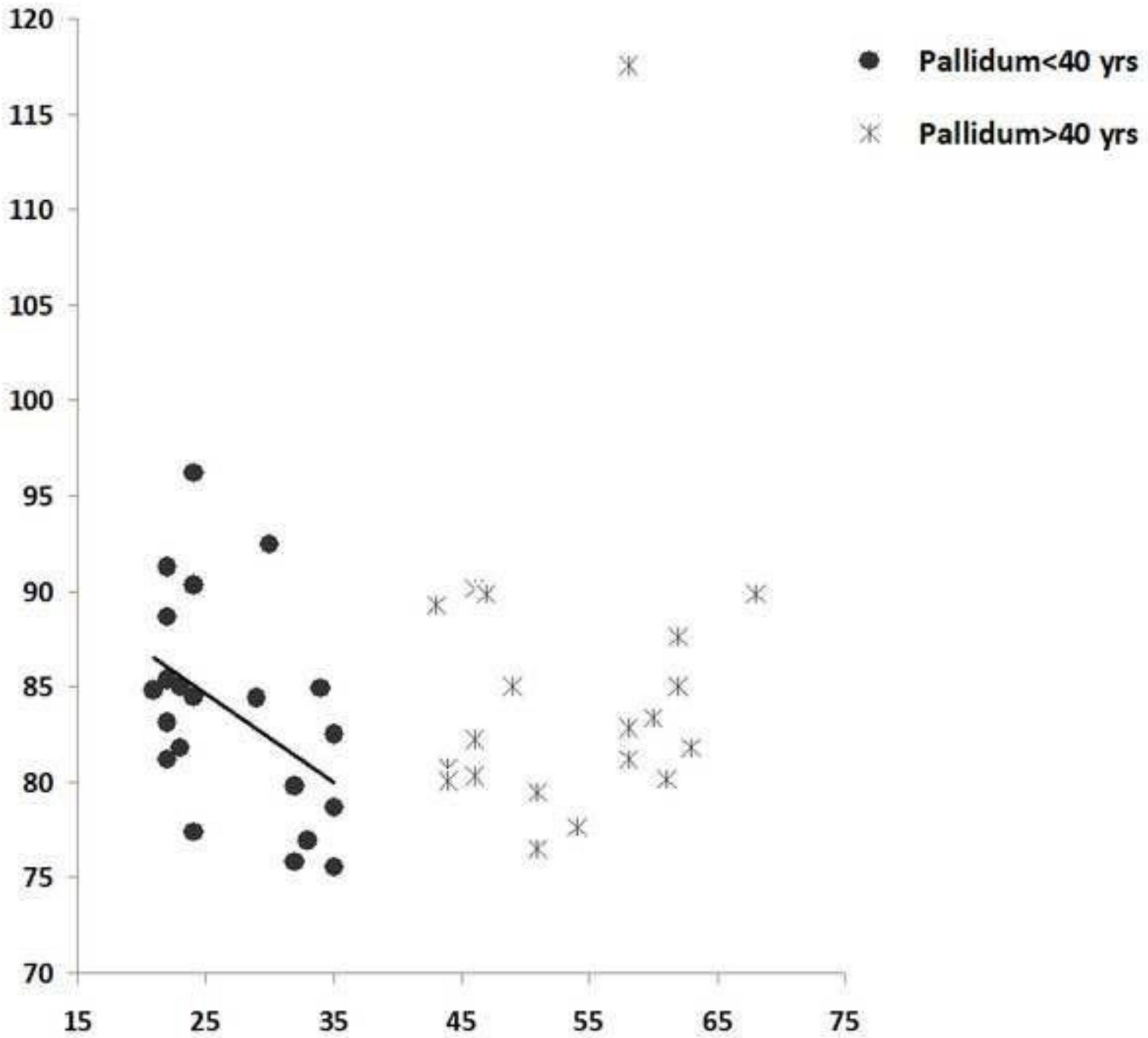

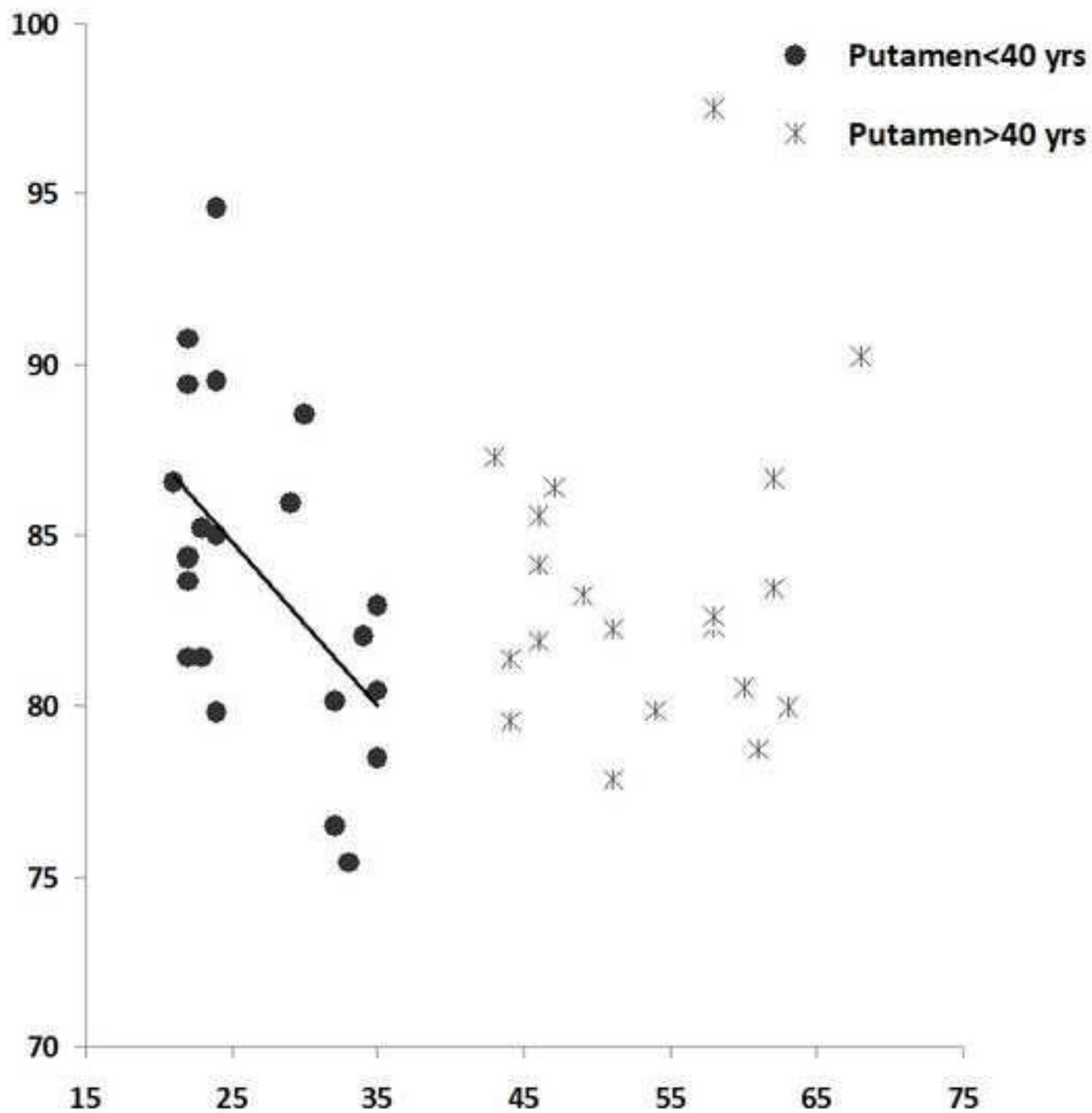

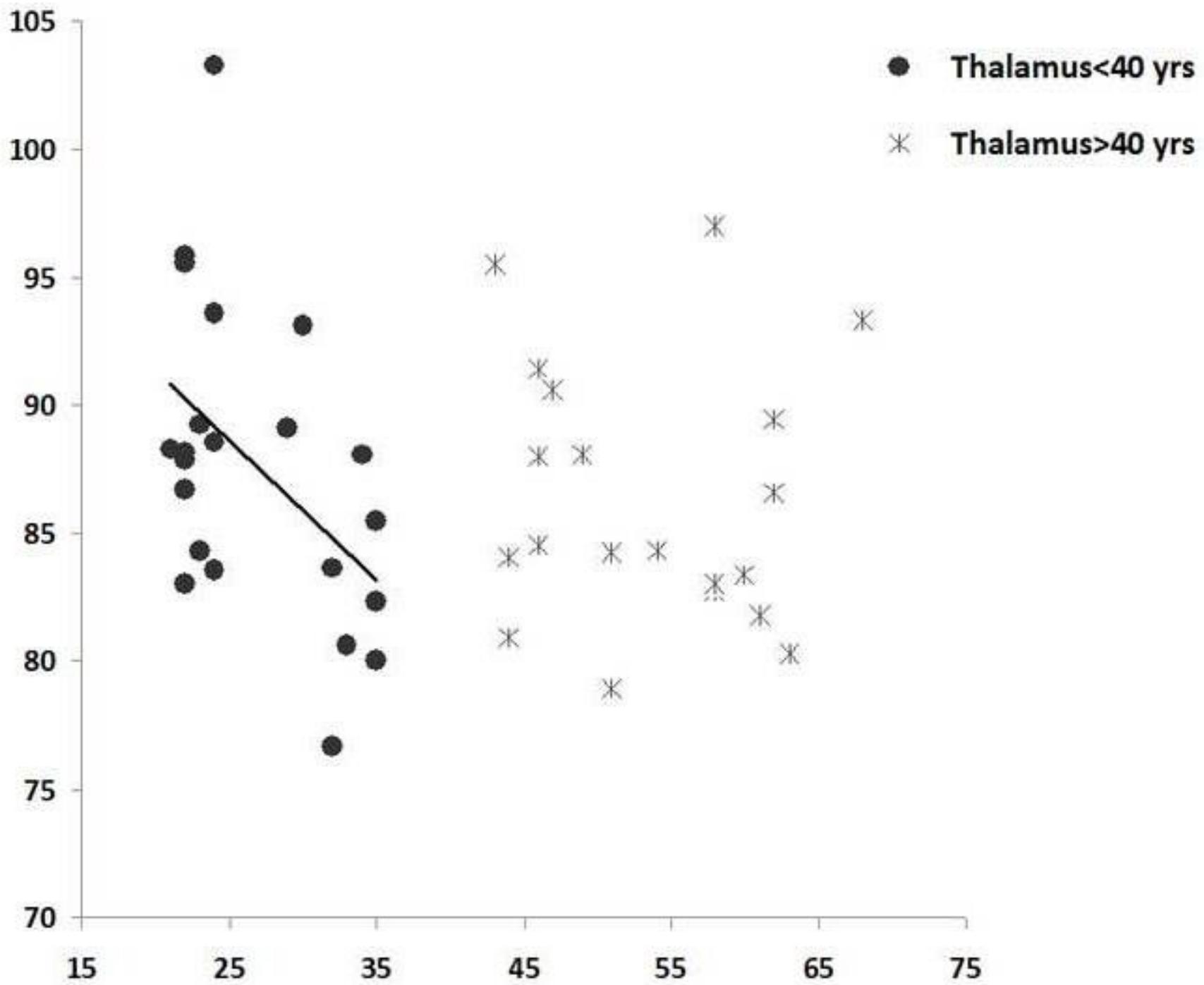

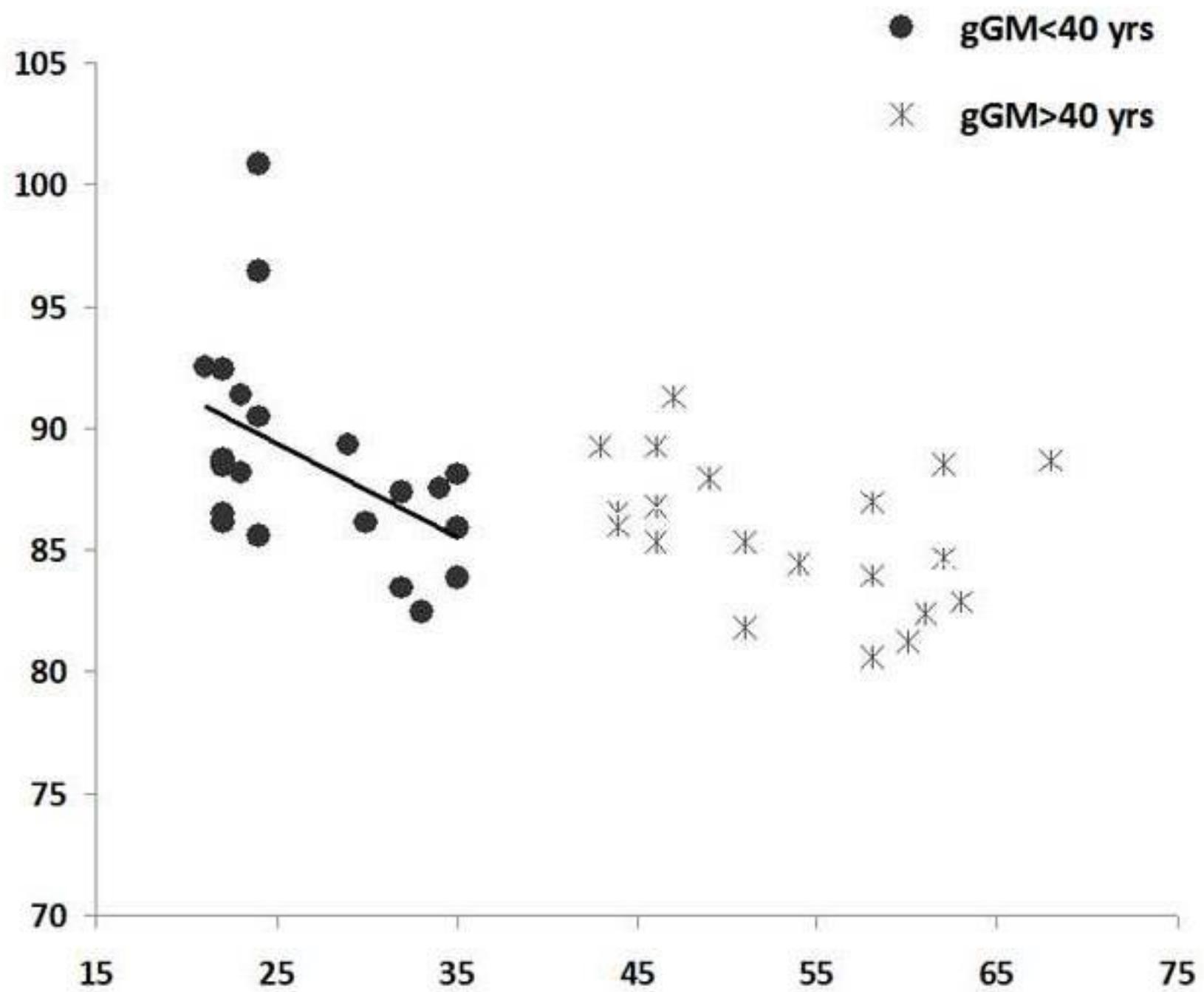

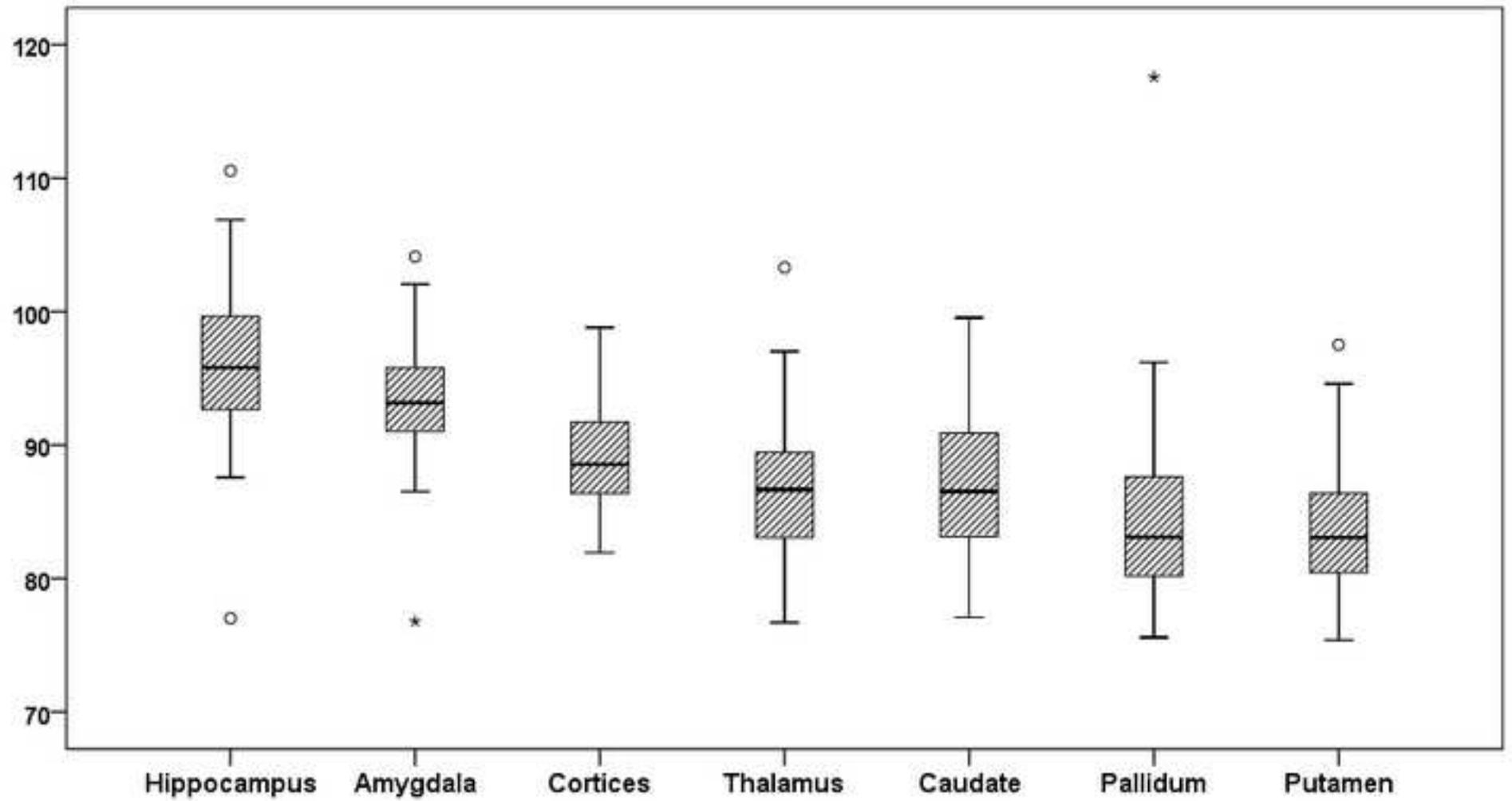

**Table 1:** Descriptive Statistics, Regression and Correlation with Age for T1ρ value of global gray matter (gGM) and selected GM structures in younger-than-40-year group

| Structure | Mean | SD | Regression coefficient with age | | Correlation | P value |
|---|---|---|---|---|---|---|
| | | | slope | R square | | |
| Amygdala* | 94.95 | 4.34 | -0.19 | 0.026 | -0.533 | 0.011 |
| Caudate* | 86.90 | 5.91 | -0.32 | 0.017 | -0.601 | 0.030 |
| Cortices** | 90.29 | 4.53 | -0.21 | 0.014 | -0.621 | 0.002 |
| Hippocampus** | 98.37 | 5.37 | -0.33 | 0.097 | -0.625 | 0.002 |
| Pallidum* | 83.81 | 5.49 | -0.15 | 0.098 | -0.449 | 0.036 |
| Putamen* | 83.93 | 4.76 | -0.23 | 0.005 | -0.537 | 0.010 |
| Thalamus* | 87.60 | 6.06 | -0.23 | 0.036 | -0.478 | 0.024 |
| gGM* | 88.66 | 4.22 | -0.39 | 0.239 | -0.489 | 0.021 |

Mean: T1ρ value in msec; SD: standard deviation, * p<0.05; p<0.005

**Table2:** Descriptive Statistics, Regression and Correlation with Age for T1ρ value of global gray matter (gGM) and selected GM structures in older-than-40-year group

| Structure | Mean | SD | Regression coefficient with age | | Correlation | P value |
|---|---|---|---|---|---|---|
| | | | slope | R square | | |
| Amygdala | 91.37 | 4.51 | -0.04 | 0.102 | -0.231 | 0.328 |
| Caudate | 86.15 | 4.43 | -0.05 | 0.145 | -0.099 | 0.678 |
| Cortices | 88.03 | 3.37 | -0.04 | 0.140 | -0.213 | 0.368 |
| Hippocampus | 93.38 | 5.73 | -0.14 | 0.171 | -0.301 | 0.198 |
| Pallidum | 85.03 | 8.72 | 0.1 | 0.019 | 0.153 | 0.519 |
| Putamen | 83.57 | 4.56 | 0.02 | 0.131 | 0.144 | 0.545 |
| Thalamus | 86.42 | 5.11 | -0.1 | 0.083 | -0.050 | 0.834 |
| gGM | 85.70 | 2.97 | -0.14 | 0.142 | -0.375 | 0.103 |

Mean: T1ρ value in msec, SD: standard deviation